\documentclass[
 aip, amsmath, amssymb, nofootinbib, reprint
 ]{revtex4-1}

\usepackage{graphicx}
\graphicspath{{./imgs_new/}}
\usepackage{dcolumn}
\usepackage{bm}
\usepackage{tcolorbox}
\usepackage[utf8]{inputenc}
\usepackage[T1]{fontenc}
\usepackage{mathptmx}
\usepackage{etoolbox}
\usepackage[export]{adjustbox}

\makeatletter
\def\@email#1#2{%
 \endgroup
 \patchcmd{\titleblock@produce}
  {\frontmatter@RRAPformat}
  {\frontmatter@RRAPformat{\produce@RRAP{*#1\href{mailto:#2}{#2}}}\frontmatter@RRAPformat}
  {}{}
}%
\makeatother
\begin{document}

\preprint{AIP/123-QED}

\title{Suppressing Modulation Instability with Reinforcement Learning}

\author{N.~Kalmykov$^{*}$}
\author{R.~Zagidullin$^{*}$}
\def\thefootnote{*}\footnotetext{These authors contributed equally to this work}
\affiliation{ 
Skolkovo Institute of Science and Technology, Moscow, Russia
}
\author{O.~Rogov}
\affiliation{ 
Skolkovo Institute of Science and Technology, Moscow, Russia
}
\affiliation{
Artificial Intelligence Research Institute (AIRI), Moscow, Russia
}
\author{S.~Rykovanov}
\affiliation{ 
Skolkovo Institute of Science and Technology, Moscow, Russia
}
\author{D. V.~Dylov$^{**}$}
\def\thefootnote{**}\footnotetext{Corresponding author: d.dylov@skol.tech}
\affiliation{ 
Skolkovo Institute of Science and Technology, Moscow, Russia
}
\affiliation{
Artificial Intelligence Research Institute (AIRI), Moscow, Russia
}

\date{\today}

\begin{abstract}
Modulation instability is a phenomenon of spontaneous pattern formation in nonlinear media, oftentimes leading to an unpredictable behaviour and a degradation of a signal of interest. 
We propose an approach based on reinforcement learning to suppress the unstable modes in the system by optimizing the parameters for the time modulation of the potential in the nonlinear system. 
We test our approach in 1D and 2D cases and propose a new class of physically-meaningful reward functions to guarantee tamed instability. 
\end{abstract}

\maketitle

Modulation Instability (MI) is one of the main challenges in nonlinear dynamical systems. The phenomenon occurs due to the coupling of unstable modes, once it overwhelms the natural dissipation of energy via diffraction or dispersion~\cite{Zakharov2009, Akhmediev1986, Sun2012, Soljacic2000, Zakharov2013}. 
The exponential growth of the unstable modes leads to Spontaneous Pattern Formation (SPF)~\cite{Burgess2007, Kip2000, Klinger2001}, oftentimes resulting in an unpredictable dynamics.
The MI patterns initially form as periodic sine-like patterns, traceable in the Fourier domain as the side bands, and involve the bifurcation of the ``stripes'' arising from the disruption of the spatial or temporal symmetry in a homogeneous initial state ~\cite{Cross1993}. 

Solutions for taming or suppressing such pattern formation processes have been sought-after in a variety of disciplines where the nonlinearity is present.
For example, in nonlinear optics, the presence of MI leads to filamentation and contributes to the generation of optical solitons~\cite{kasegawa}, making precise control and manipulation of the output of the optical system for specific applications a formidable task~\cite{Erkintalo2011}. 
In computational imaging and microscopy, MI leads to unwanted interference patterns and introduces artifacts and distortions that significantly degrade the image quality~\cite{Dylov2010, Dylov2011, Dylov2011_r}. Similarly, in the field of laser physics, MI leads to the limitations of the operating power due to significant degradation of the beam quality at high power levels~\cite{Rubenchik:10}.

Numerous studies delve into the MI control challenge. For instance, the phase-sensitive properties of MI proved useful for the harmonic seeding in passive fiber resonators~\cite{Bessin2022}. The findings demonstrated a specific initial relative phase value, crucial for the elimination of MI gain.
Also,  this problem is especially important in the field of fiber optics~\cite{Perego2022}, spanning from the pioneering conceptual works~\cite{Tai1986, kasegawa} to the modern small-noise propagators in the fiber~\cite{Kraych2019}, asymmetric spectra~\cite{Liu2021}, photonic crystals~\cite{Harvey2003} and superfluids \cite{Lagudakis2022}.
MI is also a frequent phenomenon on metasurfaces~\cite{PhysRevE.107.054212, ExtremePhysRev}, where a careful adjustment of the Rabi frequency in the control field proves efficient for suppressing the unwanted patterns~\cite{Nath2023}. 
Similarly, the study~\cite{Trombettoni2006} suggests a method for mitigating MI in Bose-Einstein condensates, relying on magnetic and optical lattice trapping. 

Interestingly, the control of the coupling potential through a spatio-temporal modulation was not studied until recently~\cite{Kumar2015, Kumar2016}. In a numerical method, the authors showed that distinct ranges of parameters of the modulation can indeed suppress MI, partially or fully. 
Such spatio-temporal modulation was also employed for controlling the output of semiconductor amplifiers~\cite{Kumar2014} and for the pump adjustment in the vertical-external-cavity surface-emitting lasers~\cite{Ahmed2015}. 
The search for the optimal values of the modulation parameters is hard to generalize analytically as the resonant factors can vary abruptly, affecting one another. 
This trait motivated us to consider modern apparatus of artificial intelligence (AI) to see if one could \textit{learn these factors} instead. 

With the widespread popularization of AI, its adoption has naturally started to influence the study of nonlinear phenomena, with the machine learning~\cite{Raissi:2018:JCP,Tang:2020} and deep learning~\cite{Lusch:2018,Raissi:2018} techniques finding their use alongside the physical models. 
Today, these methods could be split into physics-informed~\cite{PINNs} and data-driven~\cite{DDNNs}, employing a powerful arsenal of mathematical tools, such as 
Bayesian optimization~\cite{Sena2021}, neuro-fuzzy networks~\cite{Babuska:2003}, and adaptive control with reinforcement learning~\cite{Sutton2015,Yang:2014}(RL). 

The latter, despite being around since 1990s~\cite{Zomaya:1994}, is still considered an under-researched area as far as the nonlinear systems are concerned.
In RL, there is no need to define the intrinsic physical laws of the environment and there is no need to process large datasets. RL agents simply observe the state of the system and micro-adjust a controlling action until the conditions for some long-term benefit could be \textit{learnt}. These evident advantages have stimulated the recent deployment of RL in nonlinear problems, such as PDE solvers~\cite{DeepXDE, Pan2018}, coupled oscillatory ensembles~\cite{Krylov:2020}, synchronization~\cite{synchronization}, and others. As yet, however, the  control of modulation instability by reinforcement learning has not been demonstrated.  


This letter is organized as follows: First, we provide the main equation and describe the numerical methods to study the appearance of MI. Then, we overview of the RL-based technique of tuning the time modulation. Lastly, we demonstrate the successful suppression of MI by the proposed RL algorithm. We delve into the influence of nonlinearity and how the algorithm learns to negate its effects.

To describe MI and SPF, our starting point is the Complex Ginzburg-Landau Equation (CGLE), a commonly employed model for predicting the evolution of nonlinear systems~\cite{Aranson2002}:


\begin{eqnarray}
\label{eq:unmodulated}
\frac{\partial A}{\partial t} = (1 - ic)(1- |A|^2)A + (i+d) \Delta A \, ,
\end{eqnarray}
where $c$ is the coefficient of the cubic (Kerr) nonlinearity, $d$ is the diffraction coefficient, $A$ is the signal that undergoes propagation.
Restricting the range of unstable wavenumbers is known to lead to the suppression of MI~\cite{Kumar2015}. To model this, the additional spatio-temporal modulation term in Eq.~\eqref{eq:unmodulated} is used. In 2D, the modulation potential can take the following form within the CGLE:

\begin{equation}
\begin{aligned}
\partial_t A &= (1 - ic)(1 - |A|^2)A + (i + d)(\partial_{xx} A + \partial_{yy} A) \\
&\quad + 4i m (\cos(qx) + \cos(qy)) \cos(\Omega t) A \, ,
\end{aligned}  
\label{eq:modulated_1d}
\end{equation}
\noindent where $x$ and $y$ are the transverse coordinates, and the variables $m$, $q$ and $\Omega$ parameterize the modulation potential. Finding the right values of these parameters can be a challenging task in the nonlinear medium. Depending on the noise in the signal, as well as the strength of nonlinearity and diffraction, the possible solutions can qualitatively differ in the way the unstable patterns form, including the formation and the bifurcation of stripes, formation and turbulence of solitons, \textit{etc}. 
To this day, a general method for suppressing the instability in nonlinear systems is not yet developed.

\begin{figure}
 \includegraphics[width=\linewidth, center]{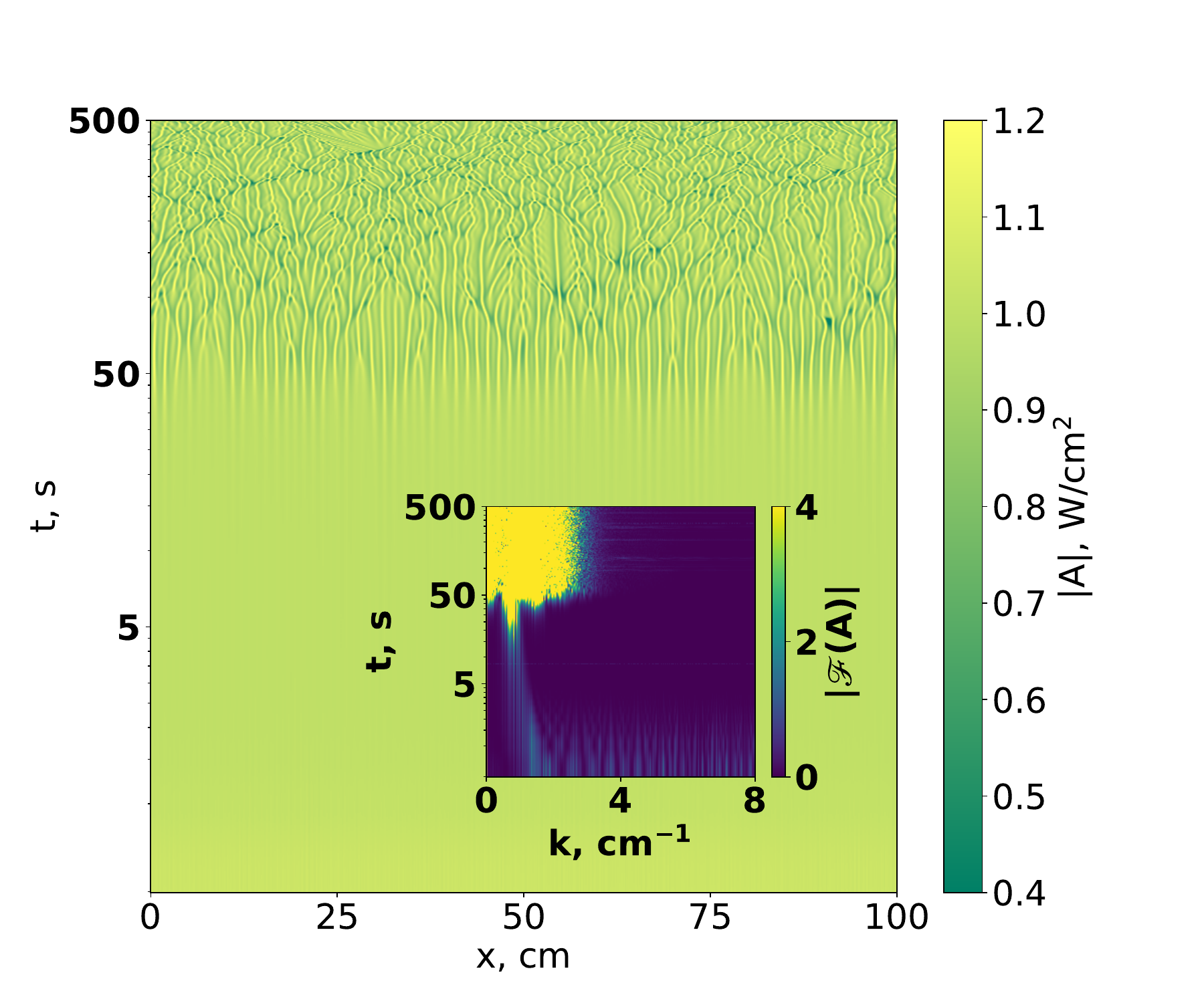}
 \caption{Numerical solution of the unmodulated CGLE with $c=0.5$, $d=0$ in 1D. Inset: spatial spectrum. The graphs are logarithmically scaled in the $t$-axis. We see generation of unstable modes due to nonlinearity. Small perturbations at every mode are present due to noise in the initial conditions. Some modes are amplified by the nonlinearity, so they can collect the energy of the propagating signal.}
 \label{fig:unmodulated}
\end{figure}

\begin{figure*}
 \includegraphics[width=\textwidth, center]{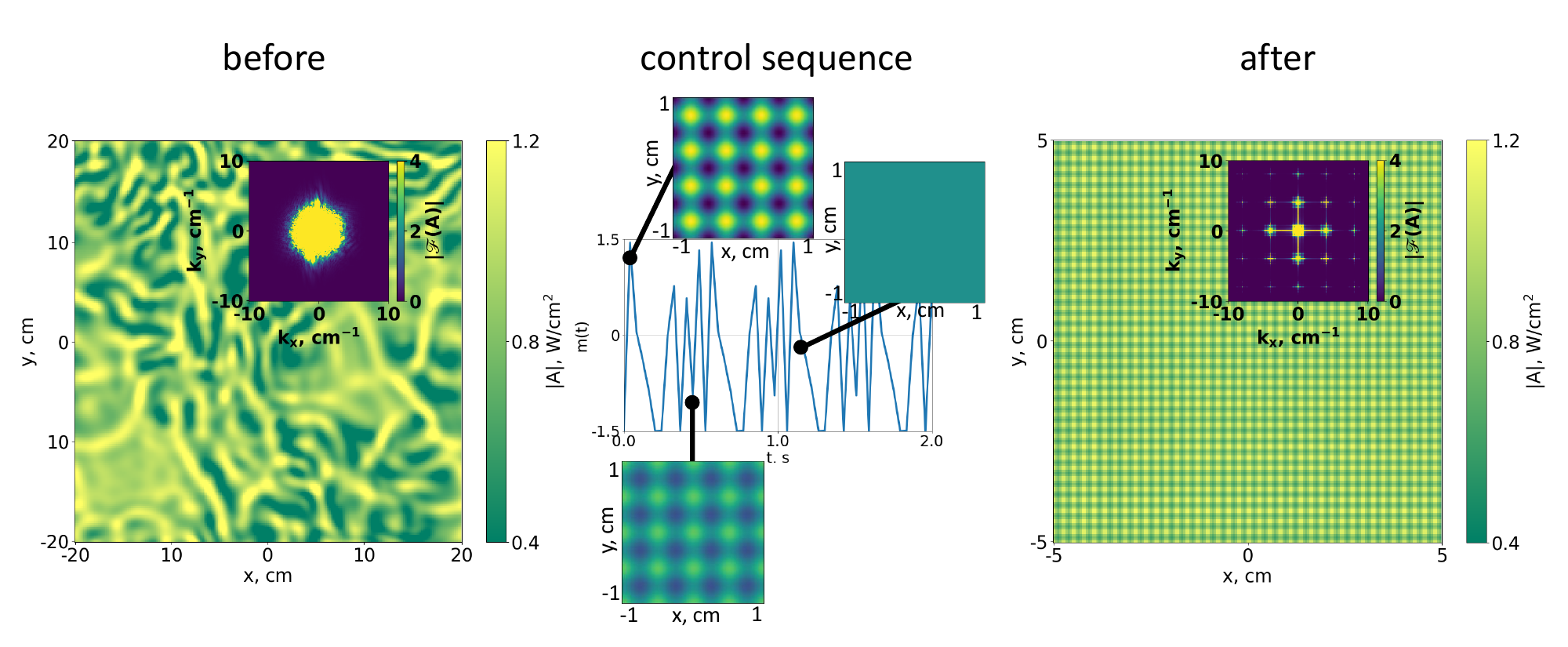}
 \caption{Description of the Q-learning algorithm training results. Left: a snapshot in time of a 2D simulation without the modulation. Unstable patterns are formed. Right: a snapshot of a 2D simulation at the end of the training. The RL-algorithm learns to modulate in time ($m(t)$) the spatial modulation of the potential (the colormaps). Insets: the snapshots in Fourier-domain. Middle: the schematic representation of the RL-control sequence. The algorithms learns to change in time the amplitude of the spatially modulated potential. If the metric $L_n$ decreases the RL-agent is rewarded according to $R_n$.}
 \label{fig2}
\end{figure*}

The Eq.~\eqref{eq:unmodulated} could be solved numerically using the split-step Fourier method. It obtains an approximate solution by assuming that in propagating the field over a small distance, the linear and nonlinear operators act independently. The linear-only part of the equation can be solved analytically in the Fourier domain. The nonlinear part (along with the potential) is solved numerically by approximating the timestep with the Taylor expansion.  It has been shown~\cite{Agraval2006} that by sandwiching a timestep of one of the operators between the two half-steps of the other can bring the accuracy of the solver to the order of 3. Example of the numerical solution of the unmodulated CGLE Eq.~\eqref{eq:unmodulated} is given in Fig.~\ref{fig:unmodulated}.

The equation can be represented as follows:\begin{equation}
\begin{aligned}
\partial_t A &= (\hat{D}+\hat{N})A \, ,
\end{aligned}
\end{equation}
where $\hat{D}$ is the linear component, $\hat{N}$ is the nonlinear component.
\begin{gather}
\begin{aligned}
\hat{N} &= (1-ic)(1-|A|^2), \quad A_{t+1} = A_t e^{\hat{N}dt} \, ,
\end{aligned}
\\
\begin{aligned}
\hat{D} &= (i + d)\partial_{xx}, \quad A_{t+1} = A_t e^{\hat{D} dt} \, .
\end{aligned}
\end{gather}

In order to solve this equation the operators have to be consequently applied  to $A$:
\begin{align}
e^{dt\hat{N}} e^{dt\hat{D}}A_t &= e^{dt\hat{N}}F^{-1}_T e^{dt \hat{D}[-i\omega]}F_TA_t \, ,
\end{align}
where $\omega$ is the frequency, $D[-i \omega]$ is obtained from the operator $\hat{D}$ by replacing $\partial_{xx}$ with $(-i\omega)^2$.

For Eq.~\eqref{eq:modulated_1d} the numerical solution is modified in accordance with the added potential:
\begin{equation}
\begin{aligned}
\hat{N} &= (1-ic)(1-|A|^2) + 4im \cos(qx)\cos(\Omega t) \, .
\end{aligned}
\end{equation}

Solving the CGLE using split-step Fourier method allows us to handle the boundary condition in Fourier space. If the signal is not padded with zeros during the Fast Fourier Transform, then the boundary condition is periodic. This was the case during our numerical experiments.

To study MI we need to perform the numerical linear stability analysis using the Floquet procedure, where the non-trivial homogeneous solutions of the CGLE are subjected to small linearly independent spatial perturbations at specific modes. The modes that cause MI can be found by analyzing the dynamics of the perturbations.

According to the Floquet theory the dynamics of a periodic solution can be analysed by constructing a monodromy matrix:

\begin{eqnarray}
\phi^{-1}(0)\phi(T) \, ,
\end{eqnarray}
where $T$ is the period of the solution, $\phi(0)$ are the linearly independent solutions to the periodic system at the beginning of the period, $\phi(T)$ -- at the end of the period.

To study the dynamics of the perturbations it suffices to keep track of the perturbed mode only ($k$ and $-k$). If the solution has additional spatial modes (e.g. the signal moves in a spatially modulated potential with modulation wavenumber $q$) the perturbation at $k$ is also coupled with modes: $q \pm k$, $2q \pm k$, $-q \pm k$, $-2q \pm k$ etc. This aids in the construction of the monodromy matrix; i.e., we are only interested in the solutions at a limited set of points. However, to construct the matrix, we need to ensure that the perturbations at mode $k$ are also linearly independent. This can be achieved numerically by using both $sine$ and $cosine$ functions as the sources of the perturbation. Additionally, the solution can be perturbed both at the real and imaginary components. Finally, we gain additional independence by perturbing the solution at any of the coupled modes mentioned above.

Once the matrix is constructed, we can find the Lyapunov exponents by taking the logarithm of the eigenvalues of the monodromy matrix:

\begin{eqnarray}
\lambda = \log \Big[ \text{Eig}(\phi^{-1}(0)\phi(T)) \Big] \, ,
\end{eqnarray}
where $\lambda$ are the Lyapunov exponents, Eig is the operation of extracting eigenvalues from a matrix.

If the Lyapunov exponents are greater than zero, it indicates the unstable solution leading to MI; otherwise, the mode is stable or is suppressed. 

\begin{figure}
  \includegraphics[width=0.4\textwidth, center]{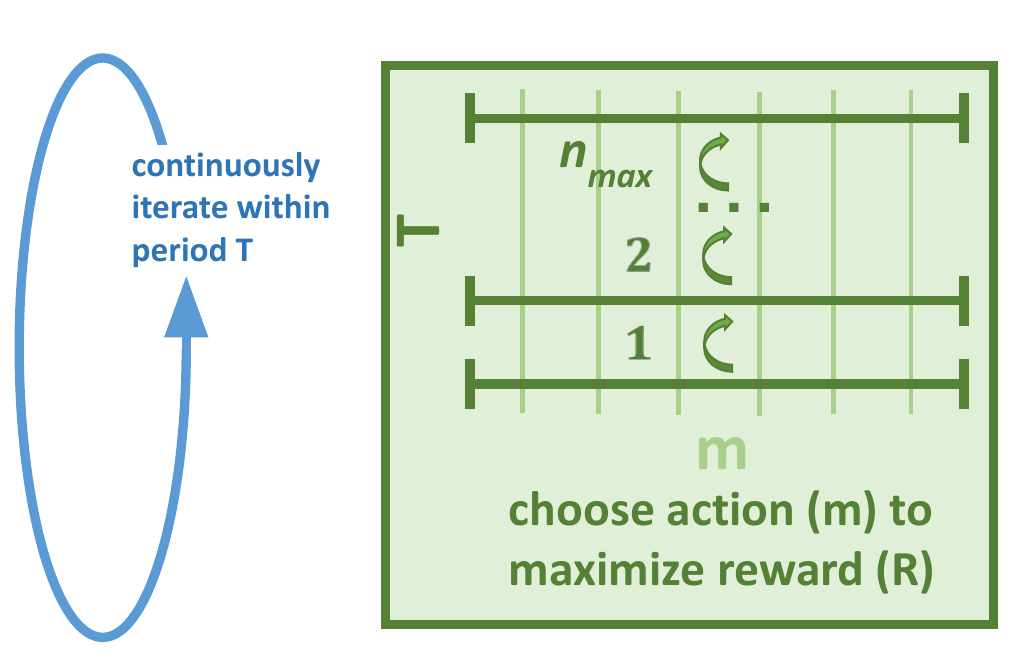}
  \caption{Schematics of Q-Learning algorithm.}
  \label{fig3}
\end{figure}

The Q-learning algorithm ~\cite{Sutton2015} relies on the construction of a table that quantifies the desirability of selecting a specific action at any state and updates each episode. Formally, this approach can be expressed with the following Temporal-Difference (TD) equation:

\begin{equation}
\begin{aligned}
Q(s, a) \leftarrow Q(s, a) + \alpha \left[ R(s, a) + \gamma \max_a Q(s', a) - Q(s, a) \right]
\end{aligned}
\end{equation}
where $Q(s, a)$ is the Q-value for the state-action ($s$-$a$) pair, $\alpha$ is the learning rate, $R(s, a)$ is the immediate reward obtained after taking an action $a$ in the state $s$, $\gamma$ is the discount factor that defines the importance of future rewards, $s'$ is the next state. Thus, the agent is supposed to learn to take correct actions by maximizing the rewards.

To automate the discovery of optimal parameters of the potential leading to full suppression of MI, we utilized an RL-based approach. Since we only cover temporal modulation, we keep the $q$ parameter fixed, while denoting the temporal modulation as an unknown function:

\begin{eqnarray}
4i \textcolor{black}{m} \cos(qx)\textcolor{black}{\cos(\Omega t)} \rightarrow 4i \cos(qx) \notag \textcolor{black}{m(t)}
\end{eqnarray}

Thus, we can consider $m(t)$ as an action space, where the agent is free to learn to correctly ``move'' the amplitude of the spatially modulated potential in time. Similar problem statement has been addressed theoretically (see \cite{Bucci2019} for a review). Here, we apply the proposed approach and develop a Q-table to control the nonlinearity that occurs in CGLE.

To differentiate the good action from bad an RL-agent needs to be given the reward function based on some metric. We use the following logic: 

1. Suppressing MI means that unstable modes cannot attract any energy of the signal. We can estimate the strength of modes as an integral in the Fourier domain:

\begin{equation}
\begin{alignedat}{2}
L_n &= \int_{k_{\text{min}}}^{k_{\text{max}}}|F_T(A_n)|dk,
\end{alignedat}
\label{L}
\end{equation}
where $A_n$ is the signal at $n$-th iteration of the numerical solution, $k_{min}$,$k_{max}$ are the boundaries of the mode range where MI occurs. So $L_n$ -- is the metric that tells us the amount of energy stored in a certain range of wavenumbers of the propagating signal.

2. The rewards need to be based on minimizing $L_n$. We propose the following reward function:

\begin{equation}
\begin{alignedat}{2}
R_n = \frac{L_{0} - L_n}{|L_{n-1} - L_n|}n^2
\end{alignedat}
\label{R}
\end{equation}

The numerator becomes negative if the total amount of energy stored in unstable modes increases compared to the initial state. The denominator motivates the agent to make sure that there are no abrupt changes in the metric values between iterations. The $n^2$ term increases the significance of the rewards and penalties in the ``late-game''

3. Lastly, we need to form the state space. To do so we first introduce parameter $T$ -- the period of the time modulation that limits the state space and, hence, the Q-table. Second, we also differentiate states based on $L_n$ using arbitrary thresholds (e.g. if $T$ consists of 13 iterations, then the agent at 1-st and 14th iterations will be placed in the same state if $L_{1}$ and $L_{14}$ are in the same threshold). 

See Fig.~\ref{fig2} for the graphic representation of the expected result once the agent learns to control the unstable modes. See Fig.~\ref{fig3} for the schematics of the inner learning loop.

\begin{figure*}
 \includegraphics[width=1.22\textwidth, center]{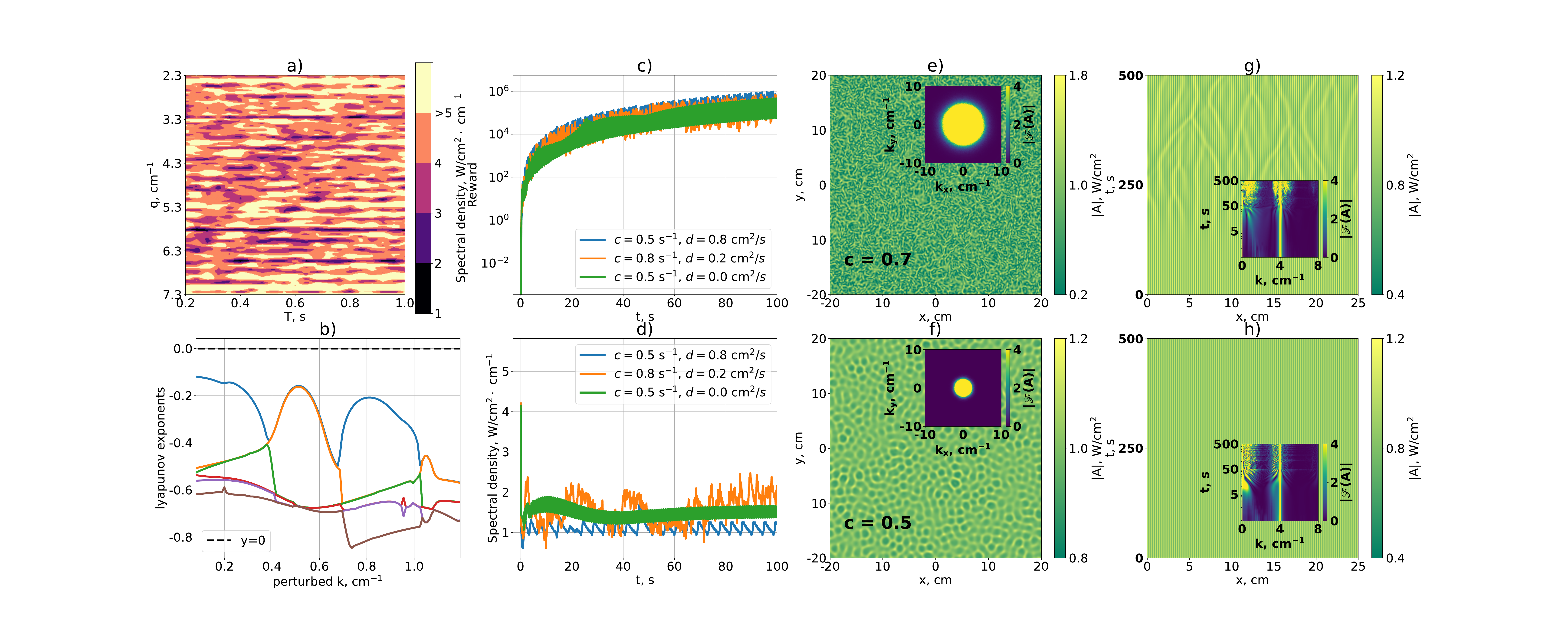}
 \caption{a) training performance of the RL-algorithm depending on parameters $q$ and $T$, b) Lyapunov exponents of time modulation provided by RL for the 1D case with $c=0.5$, $d=0.0$, c) reward (eq.~\eqref{R}) dependence on simulation time for different equation parameters, d) metric (eq.~\eqref{L}) dependence on simulation time for the same equation parameters. e) snapshot of the 2D simulation with $c=0.7$, $d=0.01$, f) the same snapshot with $c=0.5$, $d=0.01$, g) example of a failed MI suppression attempt by the trained model ($c=0.6$, $d=0.2$) in 1D simulation, h) example of a partially successful MI suppression attempt ($c=0.6$, $d=0.2$).}
 \label{fig4}
\end{figure*}

Let us describe the results for RL based time modulation for the long-term dynamics of the nonlinear system. In Fig.~\ref{fig:unmodulated}, we saw the numerical solution for the unmodulated case, which showcases the chaotic dynamics. The presence of spatial heterogeneity demonstrates the emergence of MI and corresponds to modes with positive Lyapunov exponents. It can be seen that modulation instability takes place for wavenumbers of relatively small magnitude. However, in Fig.\ref{fig2}, we see how the RL-agent (based on Q-Learning) makes the chaotic behaviour of the systems orderly.

After a sufficient number of iterations, the RL-algorithm learns to modulate the potential in time (example of a successful time modulation for 2D simulation can be seen in the middle of Fig.~\ref{fig2}). For specific values of the parameters the RL-algorithm can successfully learn to suppress MI for 1D as well as for 2D. 

Next, we study the extent of the trained models. To do so we need to see the dependence of training on the Q-table parameters, vary parameters of the governing equation, perform stability analysis and delve into the unstable mode suppression mechanics. The relevant graphical information depicting the stages of the RL related activities can be seen in Fig.~\ref{fig4}.

Spatial modulation parameter $q$ and temporal modulation period $T$ are fixed during the training. Parameter scans of those fixed parameters show us that choice of optimal values can be a crucial step in the training process (Fig.~\ref{fig4} (a)). Once the optimal values are set we can analyze the quality of the training algorithm. The RL-agent can successfully learn the ``correct'' behavior for a range of the model parameters (Fig.~\ref{fig4} (c,d)): during our experiments we were able to successfully train the nonlinearity systems for $c \in (0, 0.8]$, and $d \in [0, 1]$. The trained model can also suppress MI for a limited set values of $c$ and $d$ during inference. The same conclusion can be made when altering the levels of noise in the initial conditions. For instance if the RL-agent is trained to suppress MI for $c=0.5$ and $d=0$ then during inference the unstable modes can be dealt with for the levels of nonlinearity up to $c=0.7$ (Fig.~\ref{fig4} (e,f)). The stability analysis using the Floquet procedure shows that the unstable modes are fully suppressed (Fig.~\ref{fig4} (b)). 

In essence, the RL-based temporal modulation fights the unstable modes similar to the classical MI suppression method described in~\cite{Kumar2015}. The spatial mode present in the potential tries to pull the energy that is transferred by the signal. If succeeded the unstable modes cannot be amplified by the nonlinearity. If not, the energy is gradually transferred to the unstable modes (Fig.~\ref{fig4} (g,h) also depicts the dynamics of the (semi-)failed energy channelling control attempts).  This means that the proposed method is applicable and can replicate modern advances achieved in the field, but suffers from the same limitations as the classical approach. Therefore the RL model that we studied needs to undergo further development.

In this paper, we studied the applicability of a Q-table approach to dealing with MI. We used an RL agent to try to learn the correct time modulation needed to suppress unstable modes of a nonlinear system. The proposed approach can be used to arrive at a periodic time modulation function that can provide suppression of MI. It works well in both 1D and 2D cases.

During inference, the trained model can tame MI for a range of nonlinear system parameters: the level of noise in the initial conditions, nonlinearity, and the diffraction coefficients. However, the learning model has its own limitations, resulting on a constraint on the ranges. The Q-learning parameters need to be extended and further modified to possibly push the limit of an RL-based MI suppression. A very interesting topic for research is developing an RL-based approach applicable during the inference to qualitatively different kinds of instabilities (\textit{e.g.}, turbulence).

Further studies will be aimed at deepening the developed approach and arriving at a generalized method to control a wide range of nonlinear effects. To do so, one needs to consider more pertinent parameters, study the effect of various target metrics, and use more complex modern models. We believe that the research of RL-based instability taming methods can pave the way for a deeper understanding of MI suppression. These insights can shed light on nonintuitive potential designs, capable of stabilizing the nonlinear systems.

\begin{acknowledgments}
This work was supported by RFBR project No. 21-51-12012.
\end{acknowledgments}

\section*{Data Availability Statement}

Data sharing is not applicable to this article as no new data were created or analyzed in this study. The code used in this study is openly available at \url{github.com/RishatZagidullin/NLSE-RL}. 
\bibliography{main}

\providecommand{\noopsort}[1]{}\providecommand{\singleletter}[1]{#1}%
\begin{thebibliography}{48}%
\makeatletter
\providecommand \@ifxundefined [1]{%
 \@ifx{#1\undefined}
}%
\providecommand \@ifnum [1]{%
 \ifnum #1\expandafter \@firstoftwo
 \else \expandafter \@secondoftwo
 \fi
}%
\providecommand \@ifx [1]{%
 \ifx #1\expandafter \@firstoftwo
 \else \expandafter \@secondoftwo
 \fi
}%
\providecommand \natexlab [1]{#1}%
\providecommand \enquote  [1]{``#1''}%
\providecommand \bibnamefont  [1]{#1}%
\providecommand \bibfnamefont [1]{#1}%
\providecommand \citenamefont [1]{#1}%
\providecommand \href@noop [0]{\@secondoftwo}%
\providecommand \href [0]{\begingroup \@sanitize@url \@href}%
\providecommand \@href[1]{\@@startlink{#1}\@@href}%
\providecommand \@@href[1]{\endgroup#1\@@endlink}%
\providecommand \@sanitize@url [0]{\catcode `\\12\catcode `\$12\catcode `\&12\catcode `\#12\catcode `\^12\catcode `\_12\catcode `\%12\relax}%
\providecommand \@@startlink[1]{}%
\providecommand \@@endlink[0]{}%
\providecommand \url  [0]{\begingroup\@sanitize@url \@url }%
\providecommand \@url [1]{\endgroup\@href {#1}{\urlprefix }}%
\providecommand \urlprefix  [0]{URL }%
\providecommand \Eprint [0]{\href }%
\providecommand \doibase [0]{http://dx.doi.org/}%
\providecommand \selectlanguage [0]{\@gobble}%
\providecommand \bibinfo  [0]{\@secondoftwo}%
\providecommand \bibfield  [0]{\@secondoftwo}%
\providecommand \translation [1]{[#1]}%
\providecommand \BibitemOpen [0]{}%
\providecommand \bibitemStop [0]{}%
\providecommand \bibitemNoStop [0]{.\EOS\space}%
\providecommand \EOS [0]{\spacefactor3000\relax}%
\providecommand \BibitemShut  [1]{\csname bibitem#1\endcsname}%
\let\auto@bib@innerbib\@empty
\bibitem [{\citenamefont {Zakharov}\ and\ \citenamefont {Ostrovsky}(2009)}]{Zakharov2009}%
  \BibitemOpen
  \bibfield  {author} {\bibinfo {author} {\bibfnamefont {V.~E.}\ \bibnamefont {Zakharov}}\ and\ \bibinfo {author} {\bibfnamefont {L.~A.}\ \bibnamefont {Ostrovsky}},\ }\bibfield  {title} {\enquote {\bibinfo {title} {Modulation instability: The beginning},}\ }\href@noop {} {\bibfield  {journal} {\bibinfo  {journal} {Physica D: Nonlinear Phenomena}\ }\textbf {\bibinfo {volume} {238}},\ \bibinfo {pages} {540--548} (\bibinfo {year} {2009})}\BibitemShut {NoStop}%
\bibitem [{\citenamefont {Akhmediev}\ and\ \citenamefont {Korneev}(1986)}]{Akhmediev1986}%
  \BibitemOpen
  \bibfield  {author} {\bibinfo {author} {\bibfnamefont {N.~N.}\ \bibnamefont {Akhmediev}}\ and\ \bibinfo {author} {\bibfnamefont {V.~I.}\ \bibnamefont {Korneev}},\ }\bibfield  {title} {\enquote {\bibinfo {title} {Modulation instability and periodic solutions of the nonlinear schrodinger equation},}\ }\href@noop {} {\bibfield  {journal} {\bibinfo  {journal} {Theoretical and Mathematical Physics}\ }\textbf {\bibinfo {volume} {69}},\ \bibinfo {pages} {1089--1093} (\bibinfo {year} {1986})}\BibitemShut {NoStop}%
\bibitem [{\citenamefont {Sun}\ \emph {et~al.}(2012)\citenamefont {Sun}, \citenamefont {Waller}, \citenamefont {Dylov},\ and\ \citenamefont {Fleischer}}]{Sun2012}%
  \BibitemOpen
  \bibfield  {author} {\bibinfo {author} {\bibfnamefont {C.}~\bibnamefont {Sun}}, \bibinfo {author} {\bibfnamefont {L.}~\bibnamefont {Waller}}, \bibinfo {author} {\bibfnamefont {D.~V.}\ \bibnamefont {Dylov}}, \ and\ \bibinfo {author} {\bibfnamefont {J.~W.}\ \bibnamefont {Fleischer}},\ }\bibfield  {title} {\enquote {\bibinfo {title} {Spectral dynamics of spatially incoherent modulation instability},}\ }\href@noop {} {\bibfield  {journal} {\bibinfo  {journal} {Physical review letters}\ }\textbf {\bibinfo {volume} {108}},\ \bibinfo {pages} {263902} (\bibinfo {year} {2012})}\BibitemShut {NoStop}%
\bibitem [{\citenamefont {Soljacic}\ \emph {et~al.}(2000)\citenamefont {Soljacic}, \citenamefont {Segev}, \citenamefont {Coskun}, \citenamefont {Christodoulides},\ and\ \citenamefont {Vishwanath}}]{Soljacic2000}%
  \BibitemOpen
  \bibfield  {author} {\bibinfo {author} {\bibfnamefont {M.}~\bibnamefont {Soljacic}}, \bibinfo {author} {\bibfnamefont {M.}~\bibnamefont {Segev}}, \bibinfo {author} {\bibfnamefont {T.}~\bibnamefont {Coskun}}, \bibinfo {author} {\bibfnamefont {D.~N.}\ \bibnamefont {Christodoulides}}, \ and\ \bibinfo {author} {\bibfnamefont {A.}~\bibnamefont {Vishwanath}},\ }\bibfield  {title} {\enquote {\bibinfo {title} {Modulation instability of incoherent beams in noninstantaneous nonlinear media},}\ }\href@noop {} {\bibfield  {journal} {\bibinfo  {journal} {Physical Review Letters}\ }\textbf {\bibinfo {volume} {84}},\ \bibinfo {pages} {467} (\bibinfo {year} {2000})}\BibitemShut {NoStop}%
\bibitem [{\citenamefont {Zakharov}\ and\ \citenamefont {Gelash}(2013)}]{Zakharov2013}%
  \BibitemOpen
  \bibfield  {author} {\bibinfo {author} {\bibfnamefont {V.~E.}\ \bibnamefont {Zakharov}}\ and\ \bibinfo {author} {\bibfnamefont {A.~A.}\ \bibnamefont {Gelash}},\ }\bibfield  {title} {\enquote {\bibinfo {title} {Nonlinear stage of modulation instability},}\ }\href@noop {} {\bibfield  {journal} {\bibinfo  {journal} {Physical review letters}\ }\textbf {\bibinfo {volume} {111}},\ \bibinfo {pages} {054101} (\bibinfo {year} {2013})}\BibitemShut {NoStop}%
\bibitem [{\citenamefont {Burgess}, \citenamefont {Shimmell},\ and\ \citenamefont {Saravanamuttu}(2007)}]{Burgess2007}%
  \BibitemOpen
  \bibfield  {author} {\bibinfo {author} {\bibfnamefont {I.~B.}\ \bibnamefont {Burgess}}, \bibinfo {author} {\bibfnamefont {W.~E.}\ \bibnamefont {Shimmell}}, \ and\ \bibinfo {author} {\bibfnamefont {K.}~\bibnamefont {Saravanamuttu}},\ }\bibfield  {title} {\enquote {\bibinfo {title} {Spontaneous pattern formation due to modulation instability of incoherent white light in a photopolymerizable medium},}\ }\href@noop {} {\bibfield  {journal} {\bibinfo  {journal} {Journal of the American Chemical Society}\ }\textbf {\bibinfo {volume} {129}},\ \bibinfo {pages} {4738--4746} (\bibinfo {year} {2007})}\BibitemShut {NoStop}%
\bibitem [{\citenamefont {Kip}\ \emph {et~al.}(2000)\citenamefont {Kip}, \citenamefont {Soljacic}, \citenamefont {Segev}, \citenamefont {Eugenieva},\ and\ \citenamefont {Christodoulides}}]{Kip2000}%
  \BibitemOpen
  \bibfield  {author} {\bibinfo {author} {\bibfnamefont {D.}~\bibnamefont {Kip}}, \bibinfo {author} {\bibfnamefont {M.}~\bibnamefont {Soljacic}}, \bibinfo {author} {\bibfnamefont {M.}~\bibnamefont {Segev}}, \bibinfo {author} {\bibfnamefont {E.}~\bibnamefont {Eugenieva}}, \ and\ \bibinfo {author} {\bibfnamefont {D.~N.}\ \bibnamefont {Christodoulides}},\ }\bibfield  {title} {\enquote {\bibinfo {title} {Modulation instability and pattern formation in spatially incoherent light beams},}\ }\href@noop {} {\bibfield  {journal} {\bibinfo  {journal} {Science}\ }\textbf {\bibinfo {volume} {290}},\ \bibinfo {pages} {495--498} (\bibinfo {year} {2000})}\BibitemShut {NoStop}%
\bibitem [{\citenamefont {Klinger}, \citenamefont {Martin},\ and\ \citenamefont {Chen}(2001)}]{Klinger2001}%
  \BibitemOpen
  \bibfield  {author} {\bibinfo {author} {\bibfnamefont {J.}~\bibnamefont {Klinger}}, \bibinfo {author} {\bibfnamefont {H.}~\bibnamefont {Martin}}, \ and\ \bibinfo {author} {\bibfnamefont {Z.}~\bibnamefont {Chen}},\ }\bibfield  {title} {\enquote {\bibinfo {title} {Experiments on induced modulational instability of an incoherent optical beam},}\ }\href@noop {} {\bibfield  {journal} {\bibinfo  {journal} {Optics letters}\ }\textbf {\bibinfo {volume} {26}},\ \bibinfo {pages} {271--273} (\bibinfo {year} {2001})}\BibitemShut {NoStop}%
\bibitem [{\citenamefont {Cross}\ and\ \citenamefont {Hohenberg}(1993)}]{Cross1993}%
  \BibitemOpen
  \bibfield  {author} {\bibinfo {author} {\bibfnamefont {M.~C.}\ \bibnamefont {Cross}}\ and\ \bibinfo {author} {\bibfnamefont {P.~C.}\ \bibnamefont {Hohenberg}},\ }\bibfield  {title} {\enquote {\bibinfo {title} {Pattern formation outside of equilibrium},}\ }\href@noop {} {\bibfield  {journal} {\bibinfo  {journal} {Reviews of modern physics}\ }\textbf {\bibinfo {volume} {65}},\ \bibinfo {pages} {851} (\bibinfo {year} {1993})}\BibitemShut {NoStop}%
\bibitem [{\citenamefont {Hasegawa}(1984)}]{kasegawa}%
  \BibitemOpen
  \bibfield  {author} {\bibinfo {author} {\bibfnamefont {A.}~\bibnamefont {Hasegawa}},\ }\bibfield  {title} {\enquote {\bibinfo {title} {Generation of a train of soliton pulses by induced modulational instability in optical fibers},}\ }\href {\doibase 10.1364/OL.9.000288} {\bibfield  {journal} {\bibinfo  {journal} {Opt. Lett.}\ }\textbf {\bibinfo {volume} {9}},\ \bibinfo {pages} {288--290} (\bibinfo {year} {1984})}\BibitemShut {NoStop}%
\bibitem [{\citenamefont {Erkintalo}\ \emph {et~al.}(2011)\citenamefont {Erkintalo}, \citenamefont {Hammani}, \citenamefont {Kibler}, \citenamefont {Finot}, \citenamefont {Akhmediev}, \citenamefont {Dudley},\ and\ \citenamefont {Genty}}]{Erkintalo2011}%
  \BibitemOpen
  \bibfield  {author} {\bibinfo {author} {\bibfnamefont {M.}~\bibnamefont {Erkintalo}}, \bibinfo {author} {\bibfnamefont {K.}~\bibnamefont {Hammani}}, \bibinfo {author} {\bibfnamefont {B.}~\bibnamefont {Kibler}}, \bibinfo {author} {\bibfnamefont {C.}~\bibnamefont {Finot}}, \bibinfo {author} {\bibfnamefont {N.}~\bibnamefont {Akhmediev}}, \bibinfo {author} {\bibfnamefont {J.~M.}\ \bibnamefont {Dudley}}, \ and\ \bibinfo {author} {\bibfnamefont {G.}~\bibnamefont {Genty}},\ }\bibfield  {title} {\enquote {\bibinfo {title} {Higher-order modulation instability in nonlinear fiber optics},}\ }\href@noop {} {\bibfield  {journal} {\bibinfo  {journal} {Physical review letters}\ }\textbf {\bibinfo {volume} {107}},\ \bibinfo {pages} {253901} (\bibinfo {year} {2011})}\BibitemShut {NoStop}%
\bibitem [{\citenamefont {Dylov}\ and\ \citenamefont {Fleischer}(2010)}]{Dylov2010}%
  \BibitemOpen
  \bibfield  {author} {\bibinfo {author} {\bibfnamefont {D.~V.}\ \bibnamefont {Dylov}}\ and\ \bibinfo {author} {\bibfnamefont {J.~W.}\ \bibnamefont {Fleischer}},\ }\bibfield  {title} {\enquote {\bibinfo {title} {Nonlinear self-filtering of noisy images via dynamical stochastic resonance},}\ }\href@noop {} {\bibfield  {journal} {\bibinfo  {journal} {Nature Photonics}\ }\textbf {\bibinfo {volume} {4}},\ \bibinfo {pages} {323--328} (\bibinfo {year} {2010})}\BibitemShut {NoStop}%
\bibitem [{\citenamefont {Dylov}, \citenamefont {Waller},\ and\ \citenamefont {Fleischer}(2011{\natexlab{a}})}]{Dylov2011}%
  \BibitemOpen
  \bibfield  {author} {\bibinfo {author} {\bibfnamefont {D.~V.}\ \bibnamefont {Dylov}}, \bibinfo {author} {\bibfnamefont {L.}~\bibnamefont {Waller}}, \ and\ \bibinfo {author} {\bibfnamefont {J.~W.}\ \bibnamefont {Fleischer}},\ }\bibfield  {title} {\enquote {\bibinfo {title} {Instability-driven recovery of diffused images},}\ }\href@noop {} {\bibfield  {journal} {\bibinfo  {journal} {Optics letters}\ }\textbf {\bibinfo {volume} {36}},\ \bibinfo {pages} {3711--3713} (\bibinfo {year} {2011}{\natexlab{a}})}\BibitemShut {NoStop}%
\bibitem [{\citenamefont {Dylov}, \citenamefont {Waller},\ and\ \citenamefont {Fleischer}(2011{\natexlab{b}})}]{Dylov2011_r}%
  \BibitemOpen
  \bibfield  {author} {\bibinfo {author} {\bibfnamefont {D.~V.}\ \bibnamefont {Dylov}}, \bibinfo {author} {\bibfnamefont {L.}~\bibnamefont {Waller}}, \ and\ \bibinfo {author} {\bibfnamefont {J.~W.}\ \bibnamefont {Fleischer}},\ }\bibfield  {title} {\enquote {\bibinfo {title} {Nonlinear restoration of diffused images via seeded instability},}\ }\href@noop {} {\bibfield  {journal} {\bibinfo  {journal} {IEEE Journal of Selected Topics in Quantum Electronics}\ }\textbf {\bibinfo {volume} {18}},\ \bibinfo {pages} {916--925} (\bibinfo {year} {2011}{\natexlab{b}})}\BibitemShut {NoStop}%
\bibitem [{\citenamefont {Rubenchik}, \citenamefont {Turitsyn},\ and\ \citenamefont {Fedoruk}(2010)}]{Rubenchik:10}%
  \BibitemOpen
  \bibfield  {author} {\bibinfo {author} {\bibfnamefont {A.~M.}\ \bibnamefont {Rubenchik}}, \bibinfo {author} {\bibfnamefont {S.~K.}\ \bibnamefont {Turitsyn}}, \ and\ \bibinfo {author} {\bibfnamefont {M.~P.}\ \bibnamefont {Fedoruk}},\ }\bibfield  {title} {\enquote {\bibinfo {title} {Modulation instability in \&\#x2028;high power laser amplifiers},}\ }\href {\doibase 10.1364/OE.18.001380} {\bibfield  {journal} {\bibinfo  {journal} {Opt. Express}\ }\textbf {\bibinfo {volume} {18}},\ \bibinfo {pages} {1380--1388} (\bibinfo {year} {2010})}\BibitemShut {NoStop}%
\bibitem [{\citenamefont {Bessin}\ \emph {et~al.}(2022)\citenamefont {Bessin}, \citenamefont {Naveau1}, \citenamefont {Conforti}, \citenamefont {Kudlinski}, \citenamefont {Szriftgiser},\ and\ \citenamefont {Mussot}}]{Bessin2022}%
  \BibitemOpen
  \bibfield  {author} {\bibinfo {author} {\bibfnamefont {F.}~\bibnamefont {Bessin}}, \bibinfo {author} {\bibfnamefont {C.}~\bibnamefont {Naveau1}}, \bibinfo {author} {\bibfnamefont {M.}~\bibnamefont {Conforti}}, \bibinfo {author} {\bibfnamefont {A.}~\bibnamefont {Kudlinski}}, \bibinfo {author} {\bibfnamefont {P.}~\bibnamefont {Szriftgiser}}, \ and\ \bibinfo {author} {\bibfnamefont {A.}~\bibnamefont {Mussot}},\ }\bibfield  {title} {\enquote {\bibinfo {title} {Phase-sensitive seeded modulation instability in passive fiber resonators},}\ }\href@noop {} {\bibfield  {journal} {\bibinfo  {journal} {Communications Physics}\ }\textbf {\bibinfo {volume} {5}},\ \bibinfo {pages} {6} (\bibinfo {year} {2022})}\BibitemShut {NoStop}%
\bibitem [{\citenamefont {Perego}, \citenamefont {Bessin},\ and\ \citenamefont {Mussot}(2022)}]{Perego2022}%
  \BibitemOpen
  \bibfield  {author} {\bibinfo {author} {\bibfnamefont {A.~M.}\ \bibnamefont {Perego}}, \bibinfo {author} {\bibfnamefont {F.}~\bibnamefont {Bessin}}, \ and\ \bibinfo {author} {\bibfnamefont {A.}~\bibnamefont {Mussot}},\ }\bibfield  {title} {\enquote {\bibinfo {title} {Complexity of modulation instability},}\ }\href@noop {} {\bibfield  {journal} {\bibinfo  {journal} {Physical Review Research}\ }\textbf {\bibinfo {volume} {4}},\ \bibinfo {pages} {L022057} (\bibinfo {year} {2022})}\BibitemShut {NoStop}%
\bibitem [{\citenamefont {Tai}, \citenamefont {Hasegawa},\ and\ \citenamefont {Tomita}(1986)}]{Tai1986}%
  \BibitemOpen
  \bibfield  {author} {\bibinfo {author} {\bibfnamefont {K.}~\bibnamefont {Tai}}, \bibinfo {author} {\bibfnamefont {A.}~\bibnamefont {Hasegawa}}, \ and\ \bibinfo {author} {\bibfnamefont {A.}~\bibnamefont {Tomita}},\ }\bibfield  {title} {\enquote {\bibinfo {title} {Observation of modulational instability in optical fibers},}\ }\href@noop {} {\bibfield  {journal} {\bibinfo  {journal} {Physical review letters}\ }\textbf {\bibinfo {volume} {56}},\ \bibinfo {pages} {135} (\bibinfo {year} {1986})}\BibitemShut {NoStop}%
\bibitem [{\citenamefont {Kraych}\ \emph {et~al.}(2019)\citenamefont {Kraych}, \citenamefont {Agafontsev}, \citenamefont {Randoux},\ and\ \citenamefont {Suret}}]{Kraych2019}%
  \BibitemOpen
  \bibfield  {author} {\bibinfo {author} {\bibfnamefont {A.~E.}\ \bibnamefont {Kraych}}, \bibinfo {author} {\bibfnamefont {D.}~\bibnamefont {Agafontsev}}, \bibinfo {author} {\bibfnamefont {S.}~\bibnamefont {Randoux}}, \ and\ \bibinfo {author} {\bibfnamefont {P.}~\bibnamefont {Suret}},\ }\bibfield  {title} {\enquote {\bibinfo {title} {Statistical properties of the nonlinear stage of modulation instability in fiber optics},}\ }\href@noop {} {\bibfield  {journal} {\bibinfo  {journal} {Physical review letters}\ }\textbf {\bibinfo {volume} {123}},\ \bibinfo {pages} {093902} (\bibinfo {year} {2019})}\BibitemShut {NoStop}%
\bibitem [{\citenamefont {Liu}\ \emph {et~al.}(2021)\citenamefont {Liu}, \citenamefont {Wu}, \citenamefont {Chen}, \citenamefont {Yao},\ and\ \citenamefont {Akhmediev}}]{Liu2021}%
  \BibitemOpen
  \bibfield  {author} {\bibinfo {author} {\bibfnamefont {C.}~\bibnamefont {Liu}}, \bibinfo {author} {\bibfnamefont {Y.~H.}\ \bibnamefont {Wu}}, \bibinfo {author} {\bibfnamefont {S.~C.}\ \bibnamefont {Chen}}, \bibinfo {author} {\bibfnamefont {X.}~\bibnamefont {Yao}}, \ and\ \bibinfo {author} {\bibfnamefont {N.}~\bibnamefont {Akhmediev}},\ }\bibfield  {title} {\enquote {\bibinfo {title} {Exact analytic spectra of asymmetric modulation instability in systems with self-steepening effect},}\ }\href@noop {} {\bibfield  {journal} {\bibinfo  {journal} {Physical review letters}\ }\textbf {\bibinfo {volume} {127}},\ \bibinfo {pages} {094102} (\bibinfo {year} {2021})}\BibitemShut {NoStop}%
\bibitem [{\citenamefont {Harvey}\ \emph {et~al.}(2003)\citenamefont {Harvey}, \citenamefont {Leonhardt}, \citenamefont {Coen}, \citenamefont {Wong}, \citenamefont {Knight}, \citenamefont {Wadsworth},\ and\ \citenamefont {Russell}}]{Harvey2003}%
  \BibitemOpen
  \bibfield  {author} {\bibinfo {author} {\bibfnamefont {J.~D.}\ \bibnamefont {Harvey}}, \bibinfo {author} {\bibfnamefont {R.}~\bibnamefont {Leonhardt}}, \bibinfo {author} {\bibfnamefont {S.}~\bibnamefont {Coen}}, \bibinfo {author} {\bibfnamefont {G.~K.}\ \bibnamefont {Wong}}, \bibinfo {author} {\bibfnamefont {J.}~\bibnamefont {Knight}}, \bibinfo {author} {\bibfnamefont {W.~J.}\ \bibnamefont {Wadsworth}}, \ and\ \bibinfo {author} {\bibfnamefont {P.~S.~J.}\ \bibnamefont {Russell}},\ }\bibfield  {title} {\enquote {\bibinfo {title} {Scalar modulation instability in the normal dispersion regime by use of a photonic crystal fiber},}\ }\href@noop {} {\bibfield  {journal} {\bibinfo  {journal} {Q. J. Mech. Appl. Math.}\ }\textbf {\bibinfo {volume} {51}},\ \bibinfo {pages} {477--492} (\bibinfo {year} {2003})}\BibitemShut {NoStop}%
\bibitem [{\citenamefont {Sitnik}\ \emph {et~al.}(2022)\citenamefont {Sitnik}, \citenamefont {Alyatkin}, \citenamefont {Töpfer}, \citenamefont {Gnusov}, \citenamefont {Cookson}, \citenamefont {Sigurdsson},\ and\ \citenamefont {Lagoudakis}}]{Lagudakis2022}%
  \BibitemOpen
  \bibfield  {author} {\bibinfo {author} {\bibfnamefont {K.~A.}\ \bibnamefont {Sitnik}}, \bibinfo {author} {\bibfnamefont {S.}~\bibnamefont {Alyatkin}}, \bibinfo {author} {\bibfnamefont {J.~D.}\ \bibnamefont {Töpfer}}, \bibinfo {author} {\bibfnamefont {I.}~\bibnamefont {Gnusov}}, \bibinfo {author} {\bibfnamefont {T.}~\bibnamefont {Cookson}}, \bibinfo {author} {\bibfnamefont {H.}~\bibnamefont {Sigurdsson}}, \ and\ \bibinfo {author} {\bibfnamefont {P.}~\bibnamefont {Lagoudakis}},\ }\bibfield  {title} {\enquote {\bibinfo {title} {Spontaneous formation of time-periodic vortex cluster in nonlinear fluids of light},}\ }\href@noop {} {\bibfield  {journal} {\bibinfo  {journal} {Physical Review Letters}\ }\textbf {\bibinfo {volume} {128}},\ \bibinfo {pages} {237402} (\bibinfo {year} {2022})}\BibitemShut {NoStop}%
\bibitem [{\citenamefont {Demiquel}\ \emph {et~al.}(2023)\citenamefont {Demiquel}, \citenamefont {Achilleos}, \citenamefont {Theocharis},\ and\ \citenamefont {Tournat}}]{PhysRevE.107.054212}%
  \BibitemOpen
  \bibfield  {author} {\bibinfo {author} {\bibfnamefont {A.}~\bibnamefont {Demiquel}}, \bibinfo {author} {\bibfnamefont {V.}~\bibnamefont {Achilleos}}, \bibinfo {author} {\bibfnamefont {G.}~\bibnamefont {Theocharis}}, \ and\ \bibinfo {author} {\bibfnamefont {V.}~\bibnamefont {Tournat}},\ }\bibfield  {title} {\enquote {\bibinfo {title} {Modulation instability in nonlinear flexible mechanical metamaterials},}\ }\href@noop {} {\bibfield  {journal} {\bibinfo  {journal} {Phys. Rev. E}\ }\textbf {\bibinfo {volume} {107}},\ \bibinfo {pages} {054212} (\bibinfo {year} {2023})}\BibitemShut {NoStop}%
\bibitem [{\citenamefont {Kondratov}\ \emph {et~al.}(2016)\citenamefont {Kondratov} \emph {et~al.}}]{ExtremePhysRev}%
  \BibitemOpen
  \bibfield  {author} {\bibinfo {author} {\bibfnamefont {A.~V.}\ \bibnamefont {Kondratov}} \emph {et~al.},\ }\bibfield  {title} {\enquote {\bibinfo {title} {Extreme optical chirality of plasmonic nanohole arrays due to chiral fano resonance},}\ }\href@noop {} {\bibfield  {journal} {\bibinfo  {journal} {Phys. Rev. B}\ }\textbf {\bibinfo {volume} {93}},\ \bibinfo {pages} {195418} (\bibinfo {year} {2016})}\BibitemShut {NoStop}%
\bibitem [{\citenamefont {Nath}, \citenamefont {Mukherjee},\ and\ \citenamefont {Borgohain}(2023)}]{Nath2023}%
  \BibitemOpen
  \bibfield  {author} {\bibinfo {author} {\bibfnamefont {M.}~\bibnamefont {Nath}}, \bibinfo {author} {\bibfnamefont {R.}~\bibnamefont {Mukherjee}}, \ and\ \bibinfo {author} {\bibfnamefont {N.}~\bibnamefont {Borgohain}},\ }\bibfield  {title} {\enquote {\bibinfo {title} {Stabilization of modulation instability by control field in semiconductor quantum wells},}\ }\href@noop {} {\bibfield  {journal} {\bibinfo  {journal} {Scientific Reports}\ }\textbf {\bibinfo {volume} {13}},\ \bibinfo {pages} {7669} (\bibinfo {year} {2023})}\BibitemShut {NoStop}%
\bibitem [{\citenamefont {Trombettoni}\ \emph {et~al.}(2006)\citenamefont {Trombettoni}, \citenamefont {Kevrekidis}, \citenamefont {Nistazakis},\ and\ \citenamefont {Frantzeskakis}}]{Trombettoni2006}%
  \BibitemOpen
  \bibfield  {author} {\bibinfo {author} {\bibfnamefont {A.}~\bibnamefont {Trombettoni}}, \bibinfo {author} {\bibfnamefont {P.~G.}\ \bibnamefont {Kevrekidis}}, \bibinfo {author} {\bibfnamefont {H.~E.}\ \bibnamefont {Nistazakis}}, \ and\ \bibinfo {author} {\bibfnamefont {D.~J.}\ \bibnamefont {Frantzeskakis}},\ }\bibfield  {title} {\enquote {\bibinfo {title} {Modulational instability and its suppression for bose–einstein condensates under magnetic and optical lattice trapping},}\ }\href@noop {} {\bibfield  {journal} {\bibinfo  {journal} {Journal of Physics B: Atomic, Molecular and Optical Physics}\ }\textbf {\bibinfo {volume} {39}},\ \bibinfo {pages} {S231} (\bibinfo {year} {2006})}\BibitemShut {NoStop}%
\bibitem [{\citenamefont {Kumar}\ \emph {et~al.}(2015)\citenamefont {Kumar}, \citenamefont {Herrero}, \citenamefont {Botey},\ and\ \citenamefont {Staliunas}}]{Kumar2015}%
  \BibitemOpen
  \bibfield  {author} {\bibinfo {author} {\bibfnamefont {O.~S.}\ \bibnamefont {Kumar}}, \bibinfo {author} {\bibfnamefont {R.}~\bibnamefont {Herrero}}, \bibinfo {author} {\bibfnamefont {M.}~\bibnamefont {Botey}}, \ and\ \bibinfo {author} {\bibfnamefont {K.}~\bibnamefont {Staliunas}},\ }\bibfield  {title} {\enquote {\bibinfo {title} {Taming of modulation instability by spatio-temporal modulation of the potential},}\ }\href@noop {} {\bibfield  {journal} {\bibinfo  {journal} {Scientific Reports}\ }\textbf {\bibinfo {volume} {5}},\ \bibinfo {pages} {13268} (\bibinfo {year} {2015})}\BibitemShut {NoStop}%
\bibitem [{\citenamefont {Kumar}\ \emph {et~al.}(2016)\citenamefont {Kumar}, \citenamefont {Herrero}, \citenamefont {Botey},\ and\ \citenamefont {Staliunas}}]{Kumar2016}%
  \BibitemOpen
  \bibfield  {author} {\bibinfo {author} {\bibfnamefont {S.}~\bibnamefont {Kumar}}, \bibinfo {author} {\bibfnamefont {R.}~\bibnamefont {Herrero}}, \bibinfo {author} {\bibfnamefont {M.}~\bibnamefont {Botey}}, \ and\ \bibinfo {author} {\bibfnamefont {K.}~\bibnamefont {Staliunas}},\ }\bibfield  {title} {\enquote {\bibinfo {title} {Suppression of pattern-forming instabilities by genetic optimization},}\ }\href@noop {} {\bibfield  {journal} {\bibinfo  {journal} {Physical Review E}\ }\textbf {\bibinfo {volume} {94}},\ \bibinfo {pages} {010202} (\bibinfo {year} {2016})}\BibitemShut {NoStop}%
\bibitem [{\citenamefont {Kumar}\ \emph {et~al.}(2014)\citenamefont {Kumar}, \citenamefont {Herrero}, \citenamefont {Botey},\ and\ \citenamefont {Staliunas}}]{Kumar2014}%
  \BibitemOpen
  \bibfield  {author} {\bibinfo {author} {\bibfnamefont {S.}~\bibnamefont {Kumar}}, \bibinfo {author} {\bibfnamefont {R.}~\bibnamefont {Herrero}}, \bibinfo {author} {\bibfnamefont {M.}~\bibnamefont {Botey}}, \ and\ \bibinfo {author} {\bibfnamefont {K.}~\bibnamefont {Staliunas}},\ }\bibfield  {title} {\enquote {\bibinfo {title} {Suppression of modulation instability in broad area semiconductor amplifiers},}\ }\href@noop {} {\bibfield  {journal} {\bibinfo  {journal} {Optics letters}\ }\textbf {\bibinfo {volume} {39}},\ \bibinfo {pages} {5598--5601} (\bibinfo {year} {2014})}\BibitemShut {NoStop}%
\bibitem [{\citenamefont {Ahmed}\ \emph {et~al.}(2015)\citenamefont {Ahmed}, \citenamefont {Kumar}, \citenamefont {Herrero}, \citenamefont {Botey}, \citenamefont {Radziunas},\ and\ \citenamefont {Staliunas}}]{Ahmed2015}%
  \BibitemOpen
  \bibfield  {author} {\bibinfo {author} {\bibfnamefont {W.~W.}\ \bibnamefont {Ahmed}}, \bibinfo {author} {\bibfnamefont {S.}~\bibnamefont {Kumar}}, \bibinfo {author} {\bibfnamefont {R.}~\bibnamefont {Herrero}}, \bibinfo {author} {\bibfnamefont {M.}~\bibnamefont {Botey}}, \bibinfo {author} {\bibfnamefont {M.}~\bibnamefont {Radziunas}}, \ and\ \bibinfo {author} {\bibfnamefont {K.}~\bibnamefont {Staliunas}},\ }\bibfield  {title} {\enquote {\bibinfo {title} {Stabilization of flat-mirror vertical-external-cavity surface-emitting lasers by spatiotemporal modulation of the pump profile},}\ }\href@noop {} {\bibfield  {journal} {\bibinfo  {journal} {Physical Review A}\ }\textbf {\bibinfo {volume} {92}},\ \bibinfo {pages} {043829} (\bibinfo {year} {2015})}\BibitemShut {NoStop}%
\bibitem [{\citenamefont {Raissi}\ and\ \citenamefont {Karniadakis}(2018)}]{Raissi:2018:JCP}%
  \BibitemOpen
  \bibfield  {author} {\bibinfo {author} {\bibfnamefont {M.}~\bibnamefont {Raissi}}\ and\ \bibinfo {author} {\bibfnamefont {G.}~\bibnamefont {Karniadakis}},\ }\bibfield  {title} {\enquote {\bibinfo {title} {Hidden physics models: Machine learning of nonlinear partial differential equations},}\ }\href {\doibase https://doi.org/10.1016/j.jcp.2017.11.039} {\bibfield  {journal} {\bibinfo  {journal} {Journal of Computational Physics}\ }\textbf {\bibinfo {volume} {357}},\ \bibinfo {pages} {125--141} (\bibinfo {year} {2018})}\BibitemShut {NoStop}%
\bibitem [{\citenamefont {Tang}\ \emph {et~al.}(2020)\citenamefont {Tang}, \citenamefont {Kurths}, \citenamefont {Lin}, \citenamefont {Ott},\ and\ \citenamefont {Kocarev}}]{Tang:2020}%
  \BibitemOpen
  \bibfield  {author} {\bibinfo {author} {\bibfnamefont {Y.}~\bibnamefont {Tang}}, \bibinfo {author} {\bibfnamefont {J.}~\bibnamefont {Kurths}}, \bibinfo {author} {\bibfnamefont {W.}~\bibnamefont {Lin}}, \bibinfo {author} {\bibfnamefont {E.}~\bibnamefont {Ott}}, \ and\ \bibinfo {author} {\bibfnamefont {L.}~\bibnamefont {Kocarev}},\ }\bibfield  {title} {\enquote {\bibinfo {title} {{Introduction to Focus Issue: When machine learning meets complex systems: Networks, chaos, and nonlinear dynamics}},}\ }\href {\doibase 10.1063/5.0016505} {\bibfield  {journal} {\bibinfo  {journal} {Chaos: An Interdisciplinary Journal of Nonlinear Science}\ }\textbf {\bibinfo {volume} {30}},\ \bibinfo {pages} {063151} (\bibinfo {year} {2020})}\BibitemShut {NoStop}%
\bibitem [{\citenamefont {Lusch}, \citenamefont {Kutz},\ and\ \citenamefont {Brunton}(2018)}]{Lusch:2018}%
  \BibitemOpen
  \bibfield  {author} {\bibinfo {author} {\bibfnamefont {B.}~\bibnamefont {Lusch}}, \bibinfo {author} {\bibfnamefont {J.}~\bibnamefont {Kutz}}, \ and\ \bibinfo {author} {\bibfnamefont {S.}~\bibnamefont {Brunton}},\ }\bibfield  {title} {\enquote {\bibinfo {title} {Deep learning for universal linear embeddings of nonlinear dynamics},}\ }\href@noop {} {\bibfield  {journal} {\bibinfo  {journal} {Nature Communications}\ }\textbf {\bibinfo {volume} {9}},\ \bibinfo {pages} {4950} (\bibinfo {year} {2018})}\BibitemShut {NoStop}%
\bibitem [{\citenamefont {Raissi}(2018)}]{Raissi:2018}%
  \BibitemOpen
  \bibfield  {author} {\bibinfo {author} {\bibfnamefont {M.}~\bibnamefont {Raissi}},\ }\bibfield  {title} {\enquote {\bibinfo {title} {Deep hidden physics models: Deep learning of nonlinear partial differential equations},}\ }\href {http://jmlr.org/papers/v19/18-046.html} {\bibfield  {journal} {\bibinfo  {journal} {Journal of Machine Learning Research}\ }\textbf {\bibinfo {volume} {19}},\ \bibinfo {pages} {1--24} (\bibinfo {year} {2018})}\BibitemShut {NoStop}%
\bibitem [{\citenamefont {Cuomo}\ \emph {et~al.}(2022)\citenamefont {Cuomo}, \citenamefont {Di~Cola}, \citenamefont {Giampaolo}, \citenamefont {Rozza}, \citenamefont {Raissi},\ and\ \citenamefont {Piccialli}}]{PINNs}%
  \BibitemOpen
  \bibfield  {author} {\bibinfo {author} {\bibfnamefont {S.}~\bibnamefont {Cuomo}}, \bibinfo {author} {\bibfnamefont {V.}~\bibnamefont {Di~Cola}}, \bibinfo {author} {\bibfnamefont {F.}~\bibnamefont {Giampaolo}}, \bibinfo {author} {\bibfnamefont {G.}~\bibnamefont {Rozza}}, \bibinfo {author} {\bibfnamefont {M.}~\bibnamefont {Raissi}}, \ and\ \bibinfo {author} {\bibfnamefont {F.}~\bibnamefont {Piccialli}},\ }\bibfield  {title} {\enquote {\bibinfo {title} {Scientific machine learning through physics–informed neural networks: Where we are and what’s next},}\ }\href {\doibase 10.1007/s10915-022-01939-z} {\bibfield  {journal} {\bibinfo  {journal} {Journal of Scientific Computing}\ }\textbf {\bibinfo {volume} {92}},\ \bibinfo {pages} {88} (\bibinfo {year} {2022})}\BibitemShut {NoStop}%
\bibitem [{\citenamefont {Rudy}\ \emph {et~al.}(2017)\citenamefont {Rudy}, \citenamefont {Brunton}, \citenamefont {Proctor},\ and\ \citenamefont {Kutz}}]{DDNNs}%
  \BibitemOpen
  \bibfield  {author} {\bibinfo {author} {\bibfnamefont {S.~H.}\ \bibnamefont {Rudy}}, \bibinfo {author} {\bibfnamefont {S.~L.}\ \bibnamefont {Brunton}}, \bibinfo {author} {\bibfnamefont {J.~L.}\ \bibnamefont {Proctor}}, \ and\ \bibinfo {author} {\bibfnamefont {J.~N.}\ \bibnamefont {Kutz}},\ }\bibfield  {title} {\enquote {\bibinfo {title} {Data-driven discovery of partial differential equations},}\ }\href {\doibase 10.1126/sciadv.1602614} {\bibfield  {journal} {\bibinfo  {journal} {Science Advances}\ }\textbf {\bibinfo {volume} {3}},\ \bibinfo {pages} {e1602614} (\bibinfo {year} {2017})},\ \Eprint {http://arxiv.org/abs/https://www.science.org/doi/pdf/10.1126/sciadv.1602614} {https://www.science.org/doi/pdf/10.1126/sciadv.1602614} \BibitemShut {NoStop}%
\bibitem [{\citenamefont {Sena}\ \emph {et~al.}(2021)\citenamefont {Sena}, \citenamefont {Erkilinc}, \citenamefont {Dippon}, \citenamefont {Shariati}, \citenamefont {Emmerich}, \citenamefont {Fischer},\ and\ \citenamefont {Freund}}]{Sena2021}%
  \BibitemOpen
  \bibfield  {author} {\bibinfo {author} {\bibfnamefont {M.}~\bibnamefont {Sena}}, \bibinfo {author} {\bibfnamefont {M.~S.}\ \bibnamefont {Erkilinc}}, \bibinfo {author} {\bibfnamefont {T.}~\bibnamefont {Dippon}}, \bibinfo {author} {\bibfnamefont {B.}~\bibnamefont {Shariati}}, \bibinfo {author} {\bibfnamefont {R.}~\bibnamefont {Emmerich}}, \bibinfo {author} {\bibfnamefont {J.~K.}\ \bibnamefont {Fischer}}, \ and\ \bibinfo {author} {\bibfnamefont {R.}~\bibnamefont {Freund}},\ }\bibfield  {title} {\enquote {\bibinfo {title} {Bayesian optimization for nonlinear system identification and pre-distortion in cognitive transmitters},}\ }\href {\doibase 10.1109/JLT.2021.3083676} {\bibfield  {journal} {\bibinfo  {journal} {Journal of Lightwave Technology}\ }\textbf {\bibinfo {volume} {39}},\ \bibinfo {pages} {5008--5020} (\bibinfo {year} {2021})}\BibitemShut {NoStop}%
\bibitem [{\citenamefont {Babuška}\ and\ \citenamefont {Verbruggen}(2003)}]{Babuska:2003}%
  \BibitemOpen
  \bibfield  {author} {\bibinfo {author} {\bibfnamefont {R.}~\bibnamefont {Babuška}}\ and\ \bibinfo {author} {\bibfnamefont {H.}~\bibnamefont {Verbruggen}},\ }\bibfield  {title} {\enquote {\bibinfo {title} {Neuro-fuzzy methods for nonlinear system identification},}\ }\href {\doibase https://doi.org/10.1016/S1367-5788(03)00009-9} {\bibfield  {journal} {\bibinfo  {journal} {Annual Reviews in Control}\ }\textbf {\bibinfo {volume} {27}},\ \bibinfo {pages} {73--85} (\bibinfo {year} {2003})}\BibitemShut {NoStop}%
\bibitem [{\citenamefont {Sutton}\ and\ \citenamefont {Barto}(2015)}]{Sutton2015}%
  \BibitemOpen
  \bibfield  {author} {\bibinfo {author} {\bibfnamefont {R.~S.}\ \bibnamefont {Sutton}}\ and\ \bibinfo {author} {\bibfnamefont {A.~G.}\ \bibnamefont {Barto}},\ }\enquote {\bibinfo {title} {Reinforcement learning: An introduction},}\ \ (\bibinfo  {publisher} {MIT Press},\ \bibinfo {address} {London},\ \bibinfo {year} {2014, 2015})\ Chap.~\bibinfo {chapter} {6}, pp.\ \bibinfo {pages} {157--159},\ \bibinfo {edition} {2nd}\ ed.\BibitemShut {Stop}%
\bibitem [{\citenamefont {Xiong~Yang}\ and\ \citenamefont {Wang}(2014)}]{Yang:2014}%
  \BibitemOpen
  \bibfield  {author} {\bibinfo {author} {\bibfnamefont {D.~L.}\ \bibnamefont {Xiong~Yang}}\ and\ \bibinfo {author} {\bibfnamefont {D.}~\bibnamefont {Wang}},\ }\bibfield  {title} {\enquote {\bibinfo {title} {Reinforcement learning for adaptive optimal control of unknown continuous-time nonlinear systems with input constraints},}\ }\href {\doibase 10.1080/00207179.2013.848292} {\bibfield  {journal} {\bibinfo  {journal} {International Journal of Control}\ }\textbf {\bibinfo {volume} {87}},\ \bibinfo {pages} {553--566} (\bibinfo {year} {2014})}\BibitemShut {NoStop}%
\bibitem [{\citenamefont {Zomaya}(1994)}]{Zomaya:1994}%
  \BibitemOpen
  \bibfield  {author} {\bibinfo {author} {\bibfnamefont {A.}~\bibnamefont {Zomaya}},\ }\bibfield  {title} {\enquote {\bibinfo {title} {Reinforcement learning for the adaptive control of nonlinear systems},}\ }\href {\doibase 10.1109/21.281435} {\bibfield  {journal} {\bibinfo  {journal} {IEEE Transactions on Systems, Man, and Cybernetics}\ }\textbf {\bibinfo {volume} {24}},\ \bibinfo {pages} {357--363} (\bibinfo {year} {1994})}\BibitemShut {NoStop}%
\bibitem [{\citenamefont {Lu}\ \emph {et~al.}(2021)\citenamefont {Lu}, \citenamefont {Meng}, \citenamefont {Mao},\ and\ \citenamefont {Karniadakis}}]{DeepXDE}%
  \BibitemOpen
  \bibfield  {author} {\bibinfo {author} {\bibfnamefont {L.}~\bibnamefont {Lu}}, \bibinfo {author} {\bibfnamefont {X.}~\bibnamefont {Meng}}, \bibinfo {author} {\bibfnamefont {Z.}~\bibnamefont {Mao}}, \ and\ \bibinfo {author} {\bibfnamefont {G.~E.}\ \bibnamefont {Karniadakis}},\ }\bibfield  {title} {\enquote {\bibinfo {title} {Deepxde: A deep learning library for solving differential equations},}\ }\href {\doibase 10.1137/19M1274067} {\bibfield  {journal} {\bibinfo  {journal} {SIAM Review}\ }\textbf {\bibinfo {volume} {63}},\ \bibinfo {pages} {208--228} (\bibinfo {year} {2021})}\BibitemShut {NoStop}%
\bibitem [{\citenamefont {Pan}\ \emph {et~al.}(2018)\citenamefont {Pan}, \citenamefont {Farahmand}, \citenamefont {White}, \citenamefont {Nabi}, \citenamefont {Grover},\ and\ \citenamefont {Nikovski}}]{Pan2018}%
  \BibitemOpen
  \bibfield  {author} {\bibinfo {author} {\bibfnamefont {Y.}~\bibnamefont {Pan}}, \bibinfo {author} {\bibfnamefont {A.~M.}\ \bibnamefont {Farahmand}}, \bibinfo {author} {\bibfnamefont {M.}~\bibnamefont {White}}, \bibinfo {author} {\bibfnamefont {S.}~\bibnamefont {Nabi}}, \bibinfo {author} {\bibfnamefont {P.}~\bibnamefont {Grover}}, \ and\ \bibinfo {author} {\bibfnamefont {D.}~\bibnamefont {Nikovski}},\ }\bibfield  {title} {\enquote {\bibinfo {title} {Reinforcement learning with function-valued action spaces for partial differential equation control},}\ }\href@noop {} {\bibfield  {journal} {\bibinfo  {journal} {International Conference on Machine Learning}\ ,\ \bibinfo {pages} {3986--3995}} (\bibinfo {year} {2018})}\BibitemShut {NoStop}%
\bibitem [{\citenamefont {Krylov}\ \emph {et~al.}(2020)\citenamefont {Krylov}, \citenamefont {Combes}, \citenamefont {Laroche}, \citenamefont {Rosenblum},\ and\ \citenamefont {Dylov}}]{Krylov:2020}%
  \BibitemOpen
  \bibfield  {author} {\bibinfo {author} {\bibfnamefont {D.}~\bibnamefont {Krylov}}, \bibinfo {author} {\bibfnamefont {R.}~\bibnamefont {Combes}}, \bibinfo {author} {\bibfnamefont {R.}~\bibnamefont {Laroche}}, \bibinfo {author} {\bibfnamefont {M.}~\bibnamefont {Rosenblum}}, \ and\ \bibinfo {author} {\bibfnamefont {D.}~\bibnamefont {Dylov}},\ }\bibfield  {title} {\enquote {\bibinfo {title} {Reinforcement learning framework for deep brain stimulation study},}\ \ }(\bibinfo {year} {2020})\ pp.\ \bibinfo {pages} {2819--2826}\BibitemShut {NoStop}%
\bibitem [{\citenamefont {Weng}\ \emph {et~al.}(2019)\citenamefont {Weng}, \citenamefont {Yang}, \citenamefont {Gu}, \citenamefont {Zhang},\ and\ \citenamefont {Small}}]{synchronization}%
  \BibitemOpen
  \bibfield  {author} {\bibinfo {author} {\bibfnamefont {T.}~\bibnamefont {Weng}}, \bibinfo {author} {\bibfnamefont {H.}~\bibnamefont {Yang}}, \bibinfo {author} {\bibfnamefont {C.}~\bibnamefont {Gu}}, \bibinfo {author} {\bibfnamefont {J.}~\bibnamefont {Zhang}}, \ and\ \bibinfo {author} {\bibfnamefont {M.}~\bibnamefont {Small}},\ }\bibfield  {title} {\enquote {\bibinfo {title} {Synchronization of chaotic systems and their machine-learning models},}\ }\href {\doibase 10.1103/PhysRevE.99.042203} {\bibfield  {journal} {\bibinfo  {journal} {Phys. Rev. E}\ }\textbf {\bibinfo {volume} {99}},\ \bibinfo {pages} {042203} (\bibinfo {year} {2019})}\BibitemShut {NoStop}%
\bibitem [{\citenamefont {Aranson}\ and\ \citenamefont {Kramer}(2002)}]{Aranson2002}%
  \BibitemOpen
  \bibfield  {author} {\bibinfo {author} {\bibfnamefont {I.~S.}\ \bibnamefont {Aranson}}\ and\ \bibinfo {author} {\bibfnamefont {L.}~\bibnamefont {Kramer}},\ }\bibfield  {title} {\enquote {\bibinfo {title} {The world of the complex ginzburg-landau equation},}\ }\href@noop {} {\bibfield  {journal} {\bibinfo  {journal} {Reviews of modern physics}\ }\textbf {\bibinfo {volume} {74}},\ \bibinfo {pages} {99} (\bibinfo {year} {2002})}\BibitemShut {NoStop}%
\bibitem [{\citenamefont {Agraval}(2006)}]{Agraval2006}%
  \BibitemOpen
  \bibfield  {author} {\bibinfo {author} {\bibfnamefont {G.}~\bibnamefont {Agraval}},\ }\enquote {\bibinfo {title} {Nonlinear fiber optics},}\ \ (\bibinfo  {publisher} {Academic Press},\ \bibinfo {address} {London},\ \bibinfo {year} {2006})\ Chap.~\bibinfo {chapter} {2}, pp.\ \bibinfo {pages} {51--57},\ \bibinfo {edition} {3rd}\ ed.\BibitemShut {Stop}%
\bibitem [{\citenamefont {Bucci}\ \emph {et~al.}(2019)\citenamefont {Bucci}, \citenamefont {Semeraro}, \citenamefont {Allauzen}, \citenamefont {Wisniewski}, \citenamefont {Cordier},\ and\ \citenamefont {Mathelin}}]{Bucci2019}%
  \BibitemOpen
  \bibfield  {author} {\bibinfo {author} {\bibfnamefont {M.~A.}\ \bibnamefont {Bucci}}, \bibinfo {author} {\bibfnamefont {O.}~\bibnamefont {Semeraro}}, \bibinfo {author} {\bibfnamefont {A.}~\bibnamefont {Allauzen}}, \bibinfo {author} {\bibfnamefont {G.}~\bibnamefont {Wisniewski}}, \bibinfo {author} {\bibfnamefont {L.}~\bibnamefont {Cordier}}, \ and\ \bibinfo {author} {\bibfnamefont {L.}~\bibnamefont {Mathelin}},\ }\bibfield  {title} {\enquote {\bibinfo {title} {Control of chaotic systems by deep reinforcement learning},}\ }\href@noop {} {\bibfield  {journal} {\bibinfo  {journal} {Proceedings of the Royal Society A}\ }\textbf {\bibinfo {volume} {475}},\ \bibinfo {pages} {20190351} (\bibinfo {year} {2019})}\BibitemShut {NoStop}%
\end{thebibliography}%

\end{document}